\documentclass[epj]{svjour}
\usepackage{graphics}
\usepackage{graphicx}
\usepackage{amssymb}
\usepackage{amsfonts}
\usepackage{bm}
\begin{document}
\title{Competition between capillarity, layering and biaxiality in a confined liquid crystal}

\author{S. Varga\inst{1}\thanks{\emph{Permanent address:} Institute of Physics, University of Pannonia,
P.O. Box 158, Veszpr\'em H-8201, Hungary.}
\and Y. Mart\'{\i}nez-Rat\'on\inst{2} \and
E. Velasco\inst{3}}

\institute{Departamento de F\'{\i}sica Te\'orica de la Materia Condensada, 
Universidad Aut\'onoma de Madrid, E-28049 Madrid, Spain \and 
Grupo Interdisciplinar de Sistemas Complejos (GISC), 
Departamento de Matem\'{a}ticas,Escuela Polit\'{e}cnica Superior,
Universidad Carlos III de Madrid, Avenida de la Universidad 30, E--28911, Legan\'{e}s, Madrid, Spain
\and 
Departamento de F\'{\i}sica Te\'orica de la Materia Condensada and Instituto de Ciencia de 
Materiales Nicol\'as Cabrera, Universidad Aut\'onoma de Madrid, E-28049 Madrid, Spain
}

\date{}
\abstract{
The effect of confinement on the phase behaviour and structure of fluids made 
of biaxial hard particles (cuboids) is examined theoretically by means of Onsager 
second-order virial theory in the limit where the long particle axes are frozen in 
a mutually parallel configuration. Confinement is induced by two parallel planar 
hard walls (slit-pore geometry), with particle long axes perpendicular to the walls 
(perfect homeotropic anchoring).  In bulk, a continuous nematic-to-smectic transition 
takes place, while shape anisotropy in the (rectangular) particle cross section induces 
biaxial ordering. As a consequence, four bulk phases, uniaxial and biaxial nematic and 
smectic phases, can be stabilised as the cross-sectional aspect ratio is varied. On 
confining the fluid, the nematic-to-smectic transition is suppressed, and either 
uniaxial or biaxial phases, separated by a continuous trasition, can be present. 
Smectic ordering develops continuously from the walls for increasing particle concentration 
(in agreement with the supression of nematic-smectic second order transition at
confinement), but first-order layering transitions, involving structures with $n$ and 
$n+1$ layers, arise in the confined fluid at high concentration. Competition between layering 
and uniaxial-biaxial ordering leads to three different types of layering transitions, 
at which the two coexisting structures can be both uniaxial, one uniaxial and another 
biaxial, or both biaxial. Also, the interplay between molecular biaxiality and wall 
interactions is very subtle: while the hard wall disfavours the formation of the biaxial 
phase, biaxiality is against the layering transitions, as we have shown by comparing 
the confined phase behaviour of cylinders and cuboids.  The predictive power of Onsager 
theory is checked and confirmed by performing some calculations based on fundamental-
measure theory.
} 
\maketitle

\section{Introduction}

The behaviour of confined nematics is a problem of great fundamental and 
practical interest. Nematic materials exhibit strong responses to 
even subtle external or surface fields, which are used in various technological
applications. In general, confinement into a 
narrow planar slit pore induces competing capillary and layering transitions
at low temperature or high concentration.
Recent theoretical works have shown that these phenomena lead to complex phase 
behaviour in confined lamellar or smectic phases in three 
\cite{Binder,Ciach,delasHeras} and two \cite{M-R} dimensions. 
Classical techniques \cite{Yokoyama,Wittebrood,Musevic2} and novel atomic-force microscopy 
have been used to study capillary nematization \cite{Musevic,Musevic1} and
the presmectic regime \cite{Moreau}, but the strong smectization regime, 
where layering and commensuration effects are crucial, is yet to be 
experimentally studied.

Another fascinating topic in the field of liquid crystals is the search for
biaxial nematic phases \cite{Luckhurst} (for a recent review, see Berardi et al. 
\cite{Berardi}). 
The recent discovery of biaxial nematics in bent-core molecules \cite{biaxial,biaxial1}
has renewed the interest in biaxiality and in the
question of what the minimum molecular interaction requirements are for bulk biaxial nematic
stability. Several factors, such as hydrogen bonding, influence the stability of the biaxial ordering \cite{McGrother}.
The onset of biaxiality and the development of long-range bulk biaxial order may also 
be greatly affected by the presence of a surface,
a situation where nontrivial coupling between biaxiality and surface
interactions may occur. {Surfaces may induce biaxial phases in one-component fluids
\cite{vanRoij,vanRoij1,Dijkstra}
and in mixtures \cite{Harnau}}. In addition, confinement may induce smectic-like
stratification of the fluid, which may be coupled to biaxial ordering.
Therefore, the interplay between biaxiality and 
the capillary nematic-smectic transition, and/or layering transitions deep in 
the smectic regime, is expected to yield interesting physics, and is the 
topic of the present paper.

Many studies have been devoted to investigate how the bulk phenomenology of 
liquid crystals is affected by strong confinement and surface interactions
\cite{Sluckin0,MMTdG,Poniewierski}, 
which may promote or discourage layering and biaxiality inside the pore.
Up to now microscopic models have been scarcely used to address this issue. Elastic 
theory has been employed to discuss the anchoring energy in a biaxial nematic phase 
(see e.g. \cite{Zhang}). The phenomenological Landau-de Gennes theory has been extensively used
to study uniaxial and biaxial nematics in films under various conditions
\cite{Sheng,Sheng1,Sluckinm1,Sluckin,Gartland,Samo,Doane,Crawford,Iannacchione,Vissenberg,Ziherl1,Ziherl,Panasyuk},
and also specifically in connection with 
the biaxial transition taking place in a thin nematic layer subject to antagonistic 
easy axes in the two surfaces \cite{Palffy,Galabova,Sarlah,Zumer} (the so-called hybrid cell).
Experimental verification of this phenomenon now exists
\cite{Carbone}. The presmectic film \cite{deGennes} and fully developed confined 
smectics \cite{Kralj} were also considered.
Models dealing only with attractive particle interactions (of the Maier-Saupe type) have also 
been used for confined nematic liquid crystals 
\cite{Yamashita,Yamashita1,Yamashita2,Yamashita3}.
Theories on lattice models \cite{MMTdG,lattice2,lattice3} and computer simulations on
lattice \cite{simlatt1,simlatt2,simlatt3,simlatt4} and continuous 
\cite{sim1,sim2,sim3,sim4,sim5,sim6,sim7,Paulo3} 
models are also abundant. 
In all these studies the issue of biaxility is implicitely or explicitely considered.

\begin{figure}
\center\includegraphics[width=2.4in]{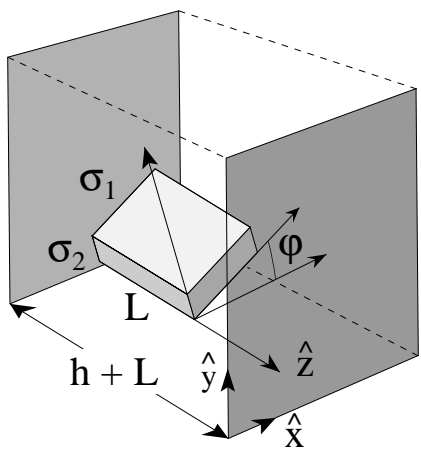}
\caption{\label{fig1}Schematic of particle shape and lengths, and geometry
used in the confined fluid.}
\end{figure}

Density-functional theory (DFT) allows us to study the properties of inhomogeneous fluids
from molecular interactions. DFT has been used to study confined nematics from a more
microscopic perspective \cite{Rull,Rull2,delasHeras3,Schmid,Reich,Paulo2}. 
A number of studies based on DFT have focused on 
confined nematic phases between symmetric \cite{Rull1,Paulo0,Paulo1,delasHeras3} and 
asymmetric \cite{Paulo3} substrates, and also confined smectic phases 
\cite{delasHeras,delasHeras1}
have been studied. All of these DFT studies considered only uniaxial, but not biaxial, particles.
{Since DFT gives a direct link between particle shape and phase biaxility,
it is a very powerful technique to investigate confined biaxial nematics and will be the
tool used in the present work.}

All mesogenic molecules are biaxial to some extent. Therefore, microscopic studies of 
idealised biaxial particle models in strong confinement and subject to
particular surface interactions seem pertinent. 
Following Vanakaras et al. \cite{Vanakaras}, we have recently investigated the bulk phase 
behaviour of a simplified athermal model for biaxial particle, consisting of hard cuboids
of lengths $\sigma_2<\sigma_1<L$, with frozen orientation in the long molecular 
axes ($z$ axis, see Fig. \ref{fig1}). We used a sophisticated free-energy 
density functional based 
on the fundamental-measure concepts \cite{Us}. The shape of these
particles effectively describes bent-core molecules, with the angle
of aperture being related to the cross-sectional aspect ratio $\kappa=\sigma_1/\sigma_2>1$ 
of the cuboids. In this model, there is always a primary nematic director $\hat{\bm z}$,
parallel to the particle long axes; but a second director may arise in the $xy$ plane,
associated with a nonuniform distribution in the azimuthal angle $\varphi$. In this case
a biaxial phase appears. In line with the findings of Vanakaras et al. \cite{Vanakaras}, based on 
Onsager theory, we found, in both the nematic and smectic regions of phase stability, 
uniaxial and biaxial phases, with a complex bulk phase diagram exhibiting continuous 
transition curves meeting at a four-phase point. 

\begin{figure}
\center\includegraphics[width=3.00in]{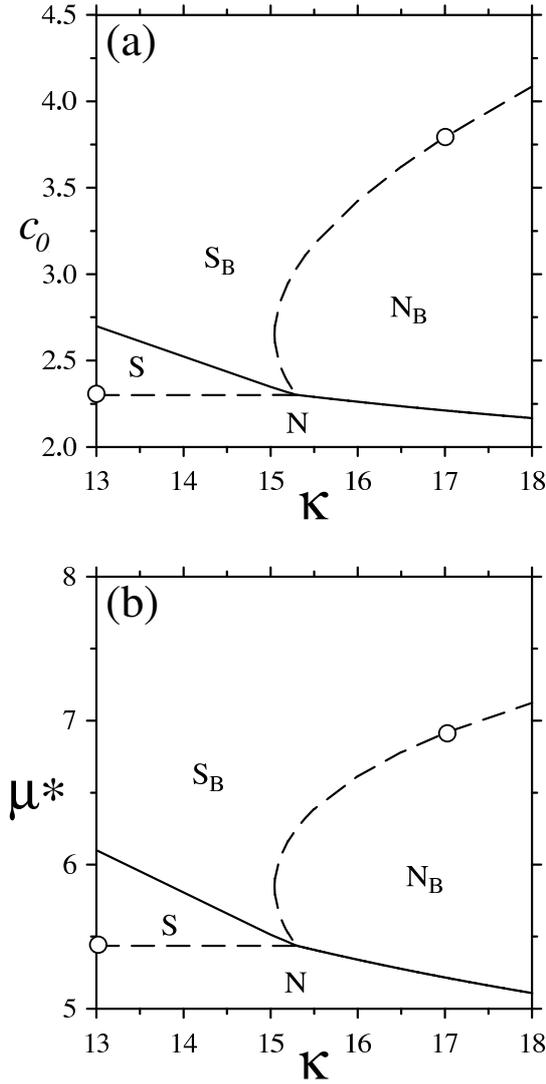}
\caption{Bulk phase diagram of hard cuboids as obtained from Onsager theory.
(a) Density $c_0$ (scaled with second virial coefficient $B_2$)
vs. cross-sectional aspect ratio $\kappa$, and (b) reduced chemical potential 
$\mu^*$ vs. $\kappa$. Curves show the boundary between the different phases.
Dashed curves indicate transitions that disappear on confinement.
All phase transitions are continuous. Labels indicate stable phases. The
corresponding bulk N-S transitions of the two confined fluids studied are indicated 
by circles.}
\label{fig2}
\end{figure}

Here we use the same Onsager theory
as in Ref. \cite{Vanakaras}, since it is much simpler than our previous
theory \cite{Us}, but predicts essentially the same qualitative behaviour. The bulk phase
diagram in the plane chemical potential $\mu$ (or scaled particle concentration $c_0$)
versus aspect ratio of the particle cross section $\kappa$, is shown in Fig. \ref{fig2} 
(see later for a more comprehensive discussion). 
The four bulk phases identified are: uniaxial nematic (N), biaxial nematic 
(N$_{\rm B}$), uniaxial smectic (S), and biaxial smectic (S$_{\rm B}$). 
Depending on the value of $\kappa$, biaxiality may arise in the nematic phase (for large 
aspect ratios) directly from the uniaxial nematic, or require as a prerequisite
the formation of smectic layers, giving rise to a biaxial smectic. 

In this work hard walls will be used to confine the fluid, promoting perfect homeotropic
anchoring (long particle axes perpendicular to the walls, Fig. \ref{fig1}). The walls do not 
couple to the azimuthal angle $\varphi$, so that the second director, whenever there is one,
can be taken to be aligned in a fixed direction, say the $\hat{\bm x}$ axis (i.e. both
walls have identical easy axes; the case with different easy axes will be the subject of
a future publication).
As will be seen later, a hard wall turns out to have a disordering effect with respect to 
biaxiality. As a consequence, biaxial phases grow in the central region of
the slab, not near the surfaces, 
and occur at higher pressure/chemical potential than in bulk. However, as common in 
adsorption problems involving hard interactions, the 
opposite effect occurs for the smectic order parameter, which is reinforced 
near the walls, so that strong layering may occur below bulk saturation.
Layering transitions, associated with commensuration effects between pore width and
smectic period, are disfavoured by biaxiality. The capillary biaxiality transition interacts
with the layering transitions in interesting ways,
giving rise to three types of layering transitions, involving two biaxial, one uniaxial and
another biaxial, or both uniaxial, coexisting structures, differing by a single layer.
Even though the biaxial bulk transition survives to confinement, the continuous 
bulk nematic-smectic transition is suppressed under confinement, so that the confined
nematic smoothly transforms into a confined smectic phase; here confined smectic phases
are identified by their propensity to undergo layering transitions at high values of
chemical potential. Each of these (infinitely large in number) first-order layering transitions 
disappears at a critical point.

Two different scenarios, corresponding to two different choices for aspect ratio, 
will be presented: $\kappa=13$ and $\kappa=17$. 
To explain these two choices, we again use Fig. \ref{fig2} 
which represents the bulk phase diagram in the reduced chemical potential $\mu^*=\beta\mu$,
or scaled particle concentration $c_0$, versus aspect ratio $\kappa$ plane. 
All bulk phase transitions are continuous and meet at a 
four-phase point. The derivatives of the transition densities with respect to the aspect 
ratio at different sides of the four-phase point are different for the four transition lines. 
The continuous nematic-to-smectic transitions, N-S and N$_{\rm B}$-S$_{\rm B}$, are suppresed under 
confinement (this is indicated by dashed curves). The uniaxial (U)-to-biaxial (B) phase 
transition (which, depending on the value of $\kappa$, is associated with either the N-N$_{\rm B}$ or S-S$_{\rm B}$ bulk
transitions) survives (this is indicated by the continuous curves), and becomes a wavy line as the wall separation
$h$ (see Fig. \ref{fig1}) is varied; its average
location is shifted to higher values of $\mu$ due to the disordering effect of the walls. 
However, even though the bulk nematic-to-smectic bulk transition disappears on confinement,
it governs the behaviour of the layering critical points, since the latter tend to the
bulk nematic-to-smectic chemical potential as the slit-pore is made wider. In one 
case considered ($\kappa=13$), the relevant bulk transition is N-S, while in the 
other ($\kappa=17$) layering transitions are governed by the N$_{\rm B}$-S$_{\rm B}$
bulk transition (open circles in Fig. \ref{fig2}). In addition, when the U-B transition is
above the bulk N-S transition ($\kappa<15.304$), layering transitions intersect with the
U-B transition curve, whereas in the opposite case the two transitions are independent.

The outline of the paper is as follows. In Section \ref{Sec2} the model system and the 
theory are presented.  The phase diagrams of bulk and confined systems and the 
equilibrium density and order parameter profiles are shown and discussed in Section 
\ref{Sec3}. In Section \ref{Sec4} we make some concluding remarks. Finally, we present 
a bifurcation analysis for the N-S, N-N$_{\rm B}$ and U-B phase transitions in the Appendix,
and an explanation for the emergence of biaxial ordering near the S-S$_B$ transition.

\section{Theory}
\label{Sec2}

We study the effect of confinement on the phase behaviour of a fluid of biaxial hard particles. 
The appropriate thermodynamic free energy for confined fluids in DFT
is the grand-canonical free-energy functional per unit area $\Omega[\rho]/A$, 
which in our case is given by
\begin{eqnarray}
\frac{\Omega[\rho]}{A}=\frac{F[\rho]}{A}-\int d{z}\int d\varphi\rho(z,\varphi)
\left[\mu-V_{\rm ext}(z,\varphi)\right],
\label{1}
\end{eqnarray}
where $\rho(z,\varphi)$
is the position- and orientation- dependent local number density, and 
$V_{\rm ext}(z,\varphi)$ is the external potential. 
In Eqn. (1) the Helmholtz free-energy
functional is the sum of ideal and excess contributions, $F[\rho]=F_{\rm id}[\rho]+F_{\rm ex}[\rho]$, 
and can be written as
\begin{eqnarray}
&&\frac{\beta F_{\rm id}[\rho]}{A}=\int dz\int d\varphi\rho(z,\varphi)
\left[\log{\left(\rho(z,\varphi)\Lambda^3\right)}-1\right],\nonumber\\\nonumber\\
&&\frac{\beta F_{\rm ex}[\rho]}{A}=-\frac{1}{2}\int dz_1\int d\varphi_1
\int dz_2\int d\varphi_2
\rho(z_1,\varphi_1)\nonumber\\\nonumber\\&&\times
\rho(z_2,\varphi_2)
\tilde{f}(z_{12},\varphi_{12}),
\label{2}
\end{eqnarray}
where $\Lambda$ is the thermal wavelength and
$\tilde{f}(z_{12},\varphi_{12})$ is the integrated Mayer function over the $x-y$ plane, with 
$z_{12}=z_1-z_2$ and $\varphi_{12}=\varphi_1-\varphi_2$. 
We set $\Lambda=1$ since this does not affect the fluid phase behaviour. Note that, while the 
ideal contribution is exact, the excess part is approximated at the level of second-order virial 
theory. Particles are confined between two parallel hard walls, the effect of which is
modelled by an external potential,
\begin{eqnarray}
\beta V_{\rm ext}(z,\varphi)=\left\{\begin{array}{cc}\infty,&z<0,\\\\
0,&0\le z\le h,\\\\\infty,&z>h,\end{array}\right.
\end{eqnarray}
where $h+L$ is the distance between the walls, and the normal to the wall is along the 
$z$ direction. This wall-particle interaction confines the centre of mass of the particles 
to stay in the finite interval $0\le z\le h$. 
Since the cross section of the cuboid is a rectangle, it is sufficient
to consider the interval $0\le\varphi\le\pi$. Due to the exclusion interaction between the wall 
and the particles, we have $\rho(z,\varphi)=0$ for $z<0$ and $z>h$. The use of frozen
orientations for the long particle axes substantially simplifies the calculations. This approximation
may be valid in a very dense mesogenic system with perfect homeotropic anchoring
(homeotropic anchoring could be achieved by means of very strong external electric/magnetic 
fields or by special surface treatments leading to specific wall-particle interactions).   
After all these assumptions, the grand canonical free energy functional per unit volume 
can be simplified
to give
\begin{eqnarray}
\frac{\Omega[\rho]}{V}=\frac{F[\rho]}{V}-\frac{\mu}{h}\int_0^h dz\rho(z),
\label{5}
\end{eqnarray}
where
\begin{eqnarray}
\rho(z)=\int_0^{\pi}d\varphi\rho(z,\varphi).
\end{eqnarray}
The ideal and excess free-energy contributions become
\begin{eqnarray}
&&\frac{\beta F_{\rm id}[\rho]}{V}=\frac{1}{h}\int_0^hdz\int_0^{\pi}d\varphi
\rho(z,\varphi)\left\{\log{\left[\rho(z,\varphi)\right]}-1\right\},\nonumber\\\nonumber\\
&&\frac{\beta F_{\rm exc}[\rho]}{V}=\frac{1}{2h}\int_0^hdz_1\int_0^{\pi}d\varphi_1
\rho(z_1,\varphi_1)
\int_{z_1-a(z_1)}^{z_1+b(z_1)}dz_2\nonumber\\\nonumber\\&&\times\int_0^{\pi}d\varphi_2
\rho(z_2,\varphi_2)A_{\rm ex}(z_{12},\varphi_{12}),
\label{6}
\end{eqnarray}
where the functions $a(z)$ and $b(z)$ are defined by
\begin{eqnarray}
a(z)=\left\{\begin{array}{cc}z,&0\le z\le L,\\\\L,&z> L,\end{array}\right.\hspace{0.6cm}
\end{eqnarray}
\begin{eqnarray}
b(z)=\left\{\begin{array}{cc}L,&z\le h-L,\\\\h-z,&h-L<z\le h.\end{array}\right.
\label{ab}
\end{eqnarray}
$A_{\rm ex}(z,\varphi)$ is the excluded area between the cross sections of two
particles. Functional minimization of 
Eqn. (\ref{5}) with respect to the local density at fixed chemical potential provides the 
equilibrium density profile 
of the fluid. The resulting integral equation for $\rho(z,\varphi)$ is
\begin{eqnarray}
&&\rho(z_1,\varphi_1)\nonumber\\\nonumber\\&&=
e^{\displaystyle \beta\mu-
\int_{z_1-a(z_1)}^{z_1+b(z_1)}\!\! dz_2\int_0^{\pi}\!\!d\varphi_2\rho(z_2,\varphi_2)A_{\rm ex}
(z_{12},\varphi_{12})}.
\label{8}
\end{eqnarray}
The inputs to the above equation are the chemical potential $\mu$ and the wall separation
$h$. The excluded area between two cuboids is given by
\begin{eqnarray}
A_{\rm ex}(z,\varphi)&=&\left[2\sigma_1\sigma_2\left(1+|\cos{\varphi}|\right)+
\left(\sigma_1^2+\sigma_2^2\right)|\sin{\varphi}|\right]\nonumber\\\nonumber\\
&&\hspace{3cm}\times\Theta(L-|z|).
\end{eqnarray}
We have solved Eqn. (\ref{8}) using two methods. One is the standard iterative procedure,
which was applied in the case of wide pores. The other is the Fourier expansion method,
used for narrow pores. The choice of method was dictated by the need to optimise
computation time, but both give numerically identical results in a wide range of pore sizes.

The properties of the bulk smectic fluid can also be calculated by solving Eqn. (\ref{8}) at fixed
chemical potential and fixed value of $h$
[in the case of the nematic, we simply evaluate (\ref{5})-(\ref{ab}) for $\rho(z,\varphi)=\rho_0=
N/V$, the mean density]. For the bulk phases, the $a(z)$ and $b(z)$ functions
have to be defined as $a(z)=b(z)=L$, and periodic boundary conditions are applied for
$\rho(z,\varphi)$ along the $z$ direction in the case of the smectic phase. The value of $h$ that
minimizes the grand potential is identified 
with the smectic period $d$, while the corresponding density distribution is the equilibrium
density distribution of the smectic phase. 
In the following, the particle length (long axis) $L$ is used as unit length, so that
the scaled pore width and the $z$ coordinate are $h^*=h/L$ and $z^*=z/L$, respectively. In the case of density,
a more convenient unit of volume is the second virial coefficient of the particles in the
uniaxial nematic phase, which is given by $B_2=LA_{\rm ex}^{(0)}$, where
$A_{\rm ex}^{(0)}=\pi^{-1}\int_0^{\pi}d\varphi A_{\rm ex}(0,\varphi)$. The scaled bulk density
will be $c_0=\rho_0 B_2$. With this choice the bulk uniaxial nematic-smectic transition 
density is independent of the geometry and shape anisotropy of the cross section (see \ref{N-S}).

Eqn. (\ref{8}) furnishes the density profile at given chemical potential for a particular
value of transverse aspect ratio. This density profile can be either uniaxial or biaxial, both 
in bulk and in the confined cases. It may also happen that more than one solution exists for
a given input. Therefore an analysis of the grand potential is necessary to find the stable 
phase and possible phase transitions. In the next section we present the bulk and confined 
phase diagrams, as well as density profiles, as obtained from the method explained above.

\begin{figure*}
\center\includegraphics[width=4.50in]{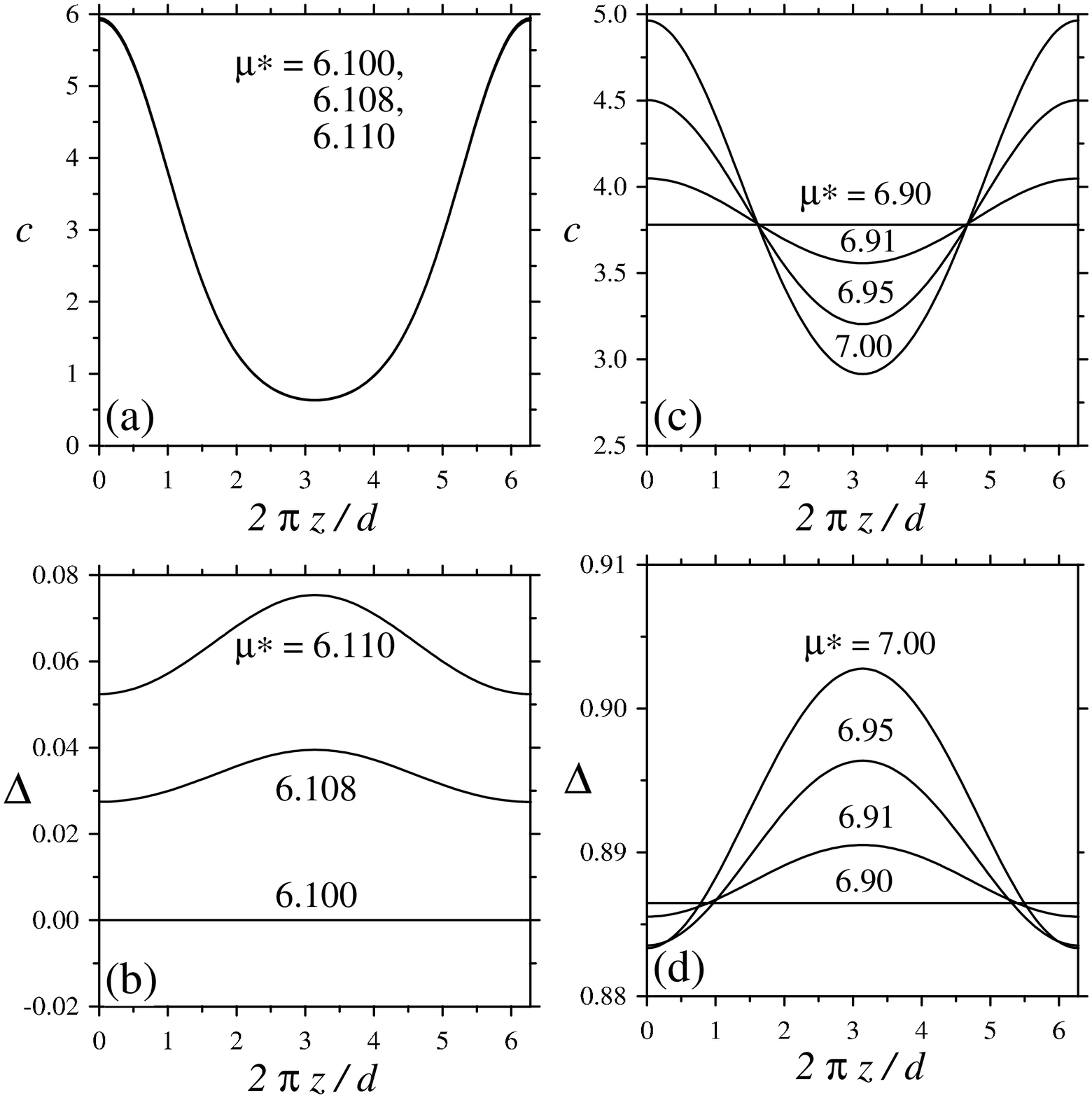}
\caption{\label{fig3}
Density profile and biaxial order parameter in the vicinity of the
S-S$_{\rm B}$ transition ((a) and (b)), and  N$_{\rm B}$-S$_{\rm B}$ transition ((c) and (d))
{for hard cuboids}. The 
aspect ratio of the particle cross section is $\kappa=13$ in (a) and (b), and $\kappa=17$
in (c) and (d). The values of the corresponding chemical potentials are shown in the 
figures.}
\end{figure*}

\section{Results and discussion}
\label{Sec3}

\subsection{Bulk behaviour}

The bulk phase diagram of the parallel hard-cuboid fluid has already been determined 
by both Onsager theory \cite{Vanakaras} and Fundamental-Measure theory (FMT) \cite{Us}. 
The main focus of those studies was to determine the effect of the cross-sectional aspect
ratio on the stability of biaxial ordering against uniaxial ordering. It was found that 
macroscopic biaxial ordering can be enhanced with increasing aspect ratio $\kappa$.
The theories agree qualitatively, in that both predict continuous uniaxial 
nematic-uniaxial smectic (N-S), uniaxial smectic-biaxial smectic (S-S$_{\rm B}$), 
uniaxial nematic-biaxial nematic (N-N$_{\rm B}$), and biaxial nematic-biaxial smectic 
(N$_{\rm B}$-S$_{\rm B}$) phase transitions, all meeting at a four-phase point located
at $(\kappa_4,\mu_4)$. However, the transition lines and the location of the four-phase
point, and therefore the onset of N$_{\rm B}$ stability, 
are different in the two theories; for example, Onsager theory predicts 
$\kappa_4=15.304$, while $\kappa_4=18.101$ with FMT. Nevertheless, it is quite surprising 
that a second-order virial theory such as Onsager's works quite well, even in the highly ordered 
smectic phase. Later in this section we show that Onsager and FMT theories also produce very 
similar density profiles in the confined system. 

\begin{figure}
\center\includegraphics[width=3.0in]{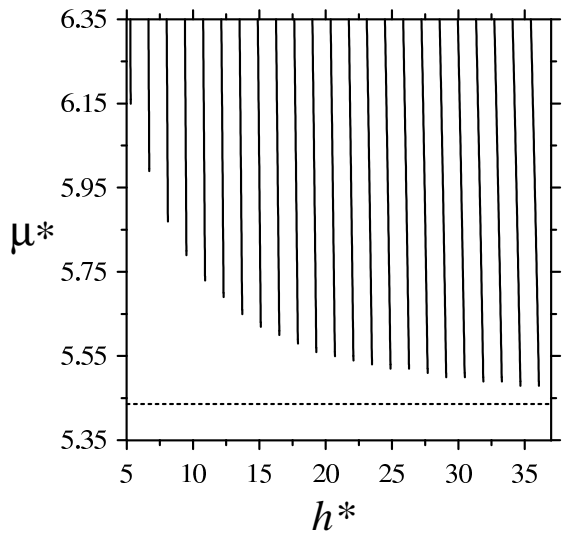}
\caption{\label{fig4}Phase diagram of confined hard cylinders. Phase boundaries of the
layering transitions in the chemical potential $\mu$ vs. pore width $h$ plane. 
The dashed line corresponds to the chemical potential of the bulk N-S transition. 
The layering transition in the widest pore examined takes place between the structures with 
26 and 27 layers.} 
\end{figure}

As a first step, we recalculated the bulk phase diagram, which is a useful reference
to discuss the properties of the confined fluid and its connection to bulk behaviour. 
Starting from a high value of $\mu$, and
performing minimisations for decreasing values, the nematic-smectic phase boundary for
a given aspect ratio $\kappa$ can be obtained when the equilibrium density distribution becomes 
independent of $z$. In addition, we performed N-S and N-N$_{\rm B}$ bifurcation analyses 
(see \ref{N-S} and \ref{N-NB} for details), the results of which agree perfectly well with those 
from the minimisations. Using $c_0$ as scaled density, the value of density at the 
boundary, $c_0^{\rm NS}$, is independent of both aspect ratio and even shape of the particle
cross section. The resulting equation for the N-S transition density is
\begin{eqnarray}
c_0^{\rm NS}=-\left[2j_0(x^*)\right]^{-1},
\end{eqnarray}
where $\pi<x^*<3\pi/2$ is the solution 
of the trascendental equation $\tan{x}=x$.
 The phase diagram is depicted in Fig. \ref{fig2}. In the calculations
that follow, we have chosen two significantly different model systems: $\kappa=13$ and $17$. As can be seen
from the figure, these two systems have different phase sequences with increasing chemical 
potential (density): N$\to$S$\to$S$_{\rm B}$ for $\kappa=13$, 
and N$\to$N$_{\rm B}\hspace{-0.1cm}\to$S$_{\rm B}$ for
the more anisotropic system ($\kappa=17$). In Fig. \ref{fig3} the local scaled density 
$c(z)$ and biaxial order parameter $\Delta(z)$, defined as
\begin{eqnarray}
&&c(z)=\frac{B_2}{\pi}\int_0^{\pi} d\varphi\rho(z,\varphi),\nonumber\\\nonumber\\
&&\Delta(z)=\frac{1}{\rho(z)}\int_0^{\pi} d\varphi\cos{2\varphi}\rho(z,\varphi),
\end{eqnarray}
are plotted as a function of $z$ along one 
smectic period. The evolution of these quantities with $\mu$ is quite interesting. For 
example, the system with $\kappa=13$ undergoes a S-S$_{\rm B}$ transition at $\mu^*=6.1$, 
but the density profile hardly changes with increasing $\mu$, while biaxial order 
develops quickly. A remarkable behaviour of the biaxial ordering is that it peaks in the 
middle of the interstitial region, which means that particles are more ordered between 
neighboring layers than inside the layers. A detailed explanation for this behaviour, 
which was already noticed in our previous FMT study \cite{Us}, is provided in Sec. \ref{S-biaxiality}. 
Regarding the N$_{\rm B}$-S$_{\rm B}$ transition of the $\kappa=17$ system [Figs. \ref{fig3}(c) and
(d)], it takes place at $\mu^*=6.9$, with $c_0=3.78$ and a biaxial order 
parameter $\Delta_0=0.886$. Both the density and the biaxial profiles change substantially 
on the smectic side. As in the previous case, the density and biaxial profiles are 
out-of-phase and, interestingly, the in-layer biaxial order decreases slightly with 
increasing chemical potential. This is evidently due to the more efficient packing in the
smectic phase.

\begin{figure}[h]
\center\includegraphics[width=3.50in]{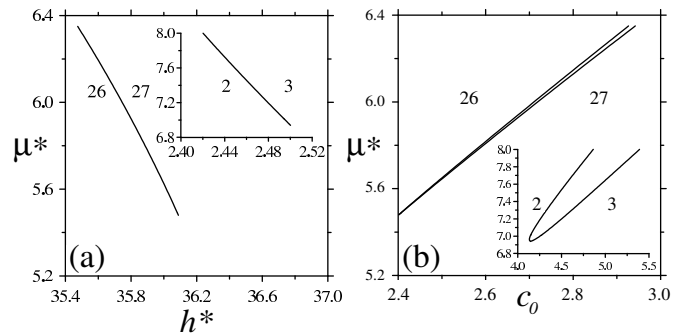}
\caption{\label{fig4a}Phase diagram of confined hard cylinders. 
(a) $\mu^*$-$h^*$ phase diagram in the region of the 26-27 layering transition.
Inset shows the 2-3 layering transition. (b) Reduced chemical potential $\mu^*$ vs. 
coexistence mean densities $c_0$ for the 26-27 layering transition.  
Inset shows the 2-3 layering transition.}
\end{figure}

\subsection{Behaviour under confinement}

\subsubsection{Hard cylinders}

We start by discussing a simpler fluid, that of parallel hard cylinders. This fluid
only undergoes a single transition in bulk, the N-S transition, and does not exhibit any 
biaxial phase (the solid phase is not considered in the present study). Due to
suppression of long-range smectic fluctuations, the N-S transition vanishes in the confined case.
However first-order layering transitions do take place. At a layering transition, two 
smectic-like structures, having $n$ and $n+1$ layers, coexist at the same pore width,
which can accommodate slightly compressed or swollen structures.
The phase diagram is shown in Fig. \ref{fig4} in the $\mu$-$h$ plane for a wide range
of pore widths. Since more and more layers can accommodate into the slit pore with increasing pore width, 
the number of layering transition is infinite in the limit $h\to\infty$. 
We can see that the layering transition 
curves are almost straight and vertical in the $\mu$-$h$ plane. However, the slope decreases
slightly with pore width. The distance between two consecutive transition curves is about 
$1.35 L$ {(with $L$ the length of the cylinders)}, and all curves terminate in lower 
critical points. The dependence of the location of these critical points on $h$
shows that the layering transitions can be stabilized with increasing pore width. 
In addition, values of the critical chemical potentials (or densities) converge to the 
bulk N-S value $\mu_{\rm NS}$ with increasing pore width. The shape of 
the layering transition curves is highlighted in Fig. \ref{fig4a}(a) by showing two extreme 
cases: the regions around the $2-3$ and $26-27$ layering transitions. The transition curves are very 
steep, but they are convex in very narrow pores and concave in wide pores. 
The dependence of the coexisting average densities with $\mu$ 
is shown in Fig. \ref{fig4a}(b), where $c_0$ is defined as $c_0=h^{-1}\int_0^h dz c(z)$.
It is clear that the density gap between coexisting 
layered structures shrinks with increasing pore width. This is due to the fact that 
the contribution from wall-particle interactions decreases relative to that from 
particle-particle interactions, i.e. the wall has less effect on the structure in the 
middle of the pore.          

\begin{figure}[h]
\center\includegraphics[width=3.0in]{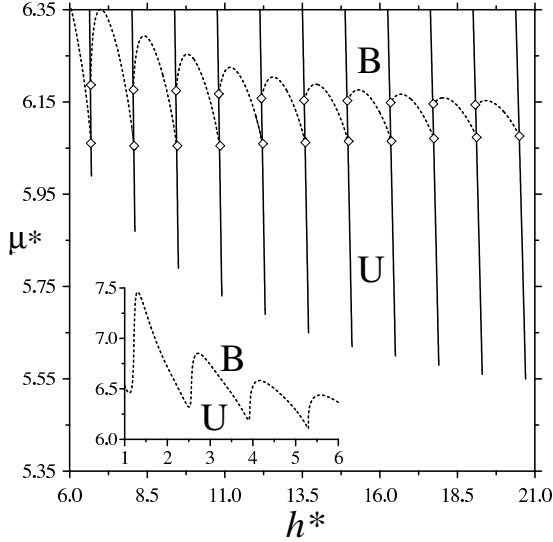}
\caption{\label{fig5}Phase diagram of confined hard cuboids in the $\mu^*$-$h^*$ plane.
Layering transitions are indicated by continuous curves, while the uniaxial-biaxial phase 
is indicated by a dashed curve. Inset shows the uniaxial-biaxial phase transition for
narrow pores. U and B denote uniaxial and biaxial phases, respectively. Diamonds indicate
the intersection between the uniaxial-biaxial and layering transitions. 
The value of the cross-sectional aspect ratio is $\kappa=13$.}
\end{figure}

\subsubsection{Hard cuboids}

In the case of confined hard cuboids again there is no N-S transition. However, by contrast with the
previous case, there is a
uniaxial-to-biaxial (U-B) phase transition. Therefore only one type of (continuous) ordering 
transition can take place. The phase diagram of confined cuboids is shown in Fig. \ref{fig5}.
While the layering transition curves are obtained by searching for two solutions of the
Euler-Lagrange equation, Eqn. (\ref{8}), with equal grand potentials, the U-B phase boundary 
was obtained from the numerical solution of the set of bifurcation equations derived in \ref{U-B} 
[see Eqns. (\ref{A8}), (\ref{A10}) and (\ref{A11})]. 
The U-B transition line $\mu_{\rm UB}(h)$ is an oscillatory function of 
pore width that asymptotically tends to the bulk S-S$_{\rm B}$ transition, $\mu^*_{{\rm SS}_B}=6.1$, with 
increasing pore width. This means that the wall separation strongly influences the 
biaxial ordering and, except for some specific intermediate ranges of pore width, confinement 
destabilizes biaxial ordering in general. For $h\gtrsim 6.6$ the curve $\mu_{\rm UB}(h)$
interacts with the layering-transition structure and becomes discontinuous. In this case
biaxiality changes the structure of the two coexisting phases at a layering transition. With 
increasing $\mu$ along a particular layering transition, first the phase at left of the 
transition curve becomes biaxial, and then both phases show biaxial order. Therefore
three types of layering transitions (U-U, U-B or B-B) can emerge in the pore. 

\begin{figure*}
\center\includegraphics[width=4.35in]{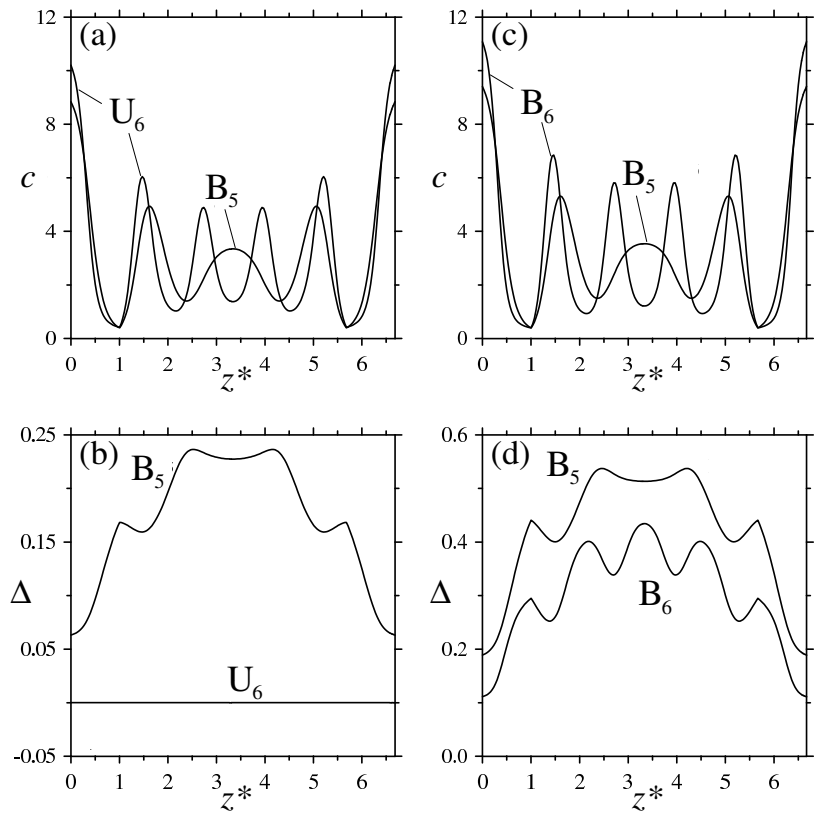}
\caption{\label{fig6} Density and order-parameter profiles of coexisting structures with
5 and 6 layers. Panels (a) and (b) show profiles at the U-B layering transition with $\mu^*=6.1$, 
while panels (c) and (d) correspond to those at the B-B transition with $\mu^*=6.3$. 
U$_5$ and U$_6$ denote uniaxial phases with 5 and 6 layers, respectively, while
B$_5$ and B$_6$ pertain to biaxial phases. The cross-sectional aspect ratio is $\kappa=13$.}
\end{figure*}

The density and order parameter profiles for two coexisting structures of types U-B and
B-B at the 5-6 layering transition are illustrated in Fig. \ref{fig6}. Due to 
commensuration effects, the density distribution of 5 layers is always less peaked than that 
of 6 layers. In the 5-layer structure, the pore width is a bit too spacious for the rods, 
while for the 6-layer structure it is too narrow. This is the case for the two coexisting 
structures at the general $n$-$n+1$ transition. Similar to the bulk biaxial smectic phase 
[see Fig. \ref{fig3}], particles are less ordered orientationally at the layers 
than in the interstitial regions. As the wall reduces biaxial order and therefore
destabilizes the U-B phase transition, the biaxial order parameter is lowest at contact with 
the wall. 

\begin{figure*}
\center\includegraphics[width=4.35in]{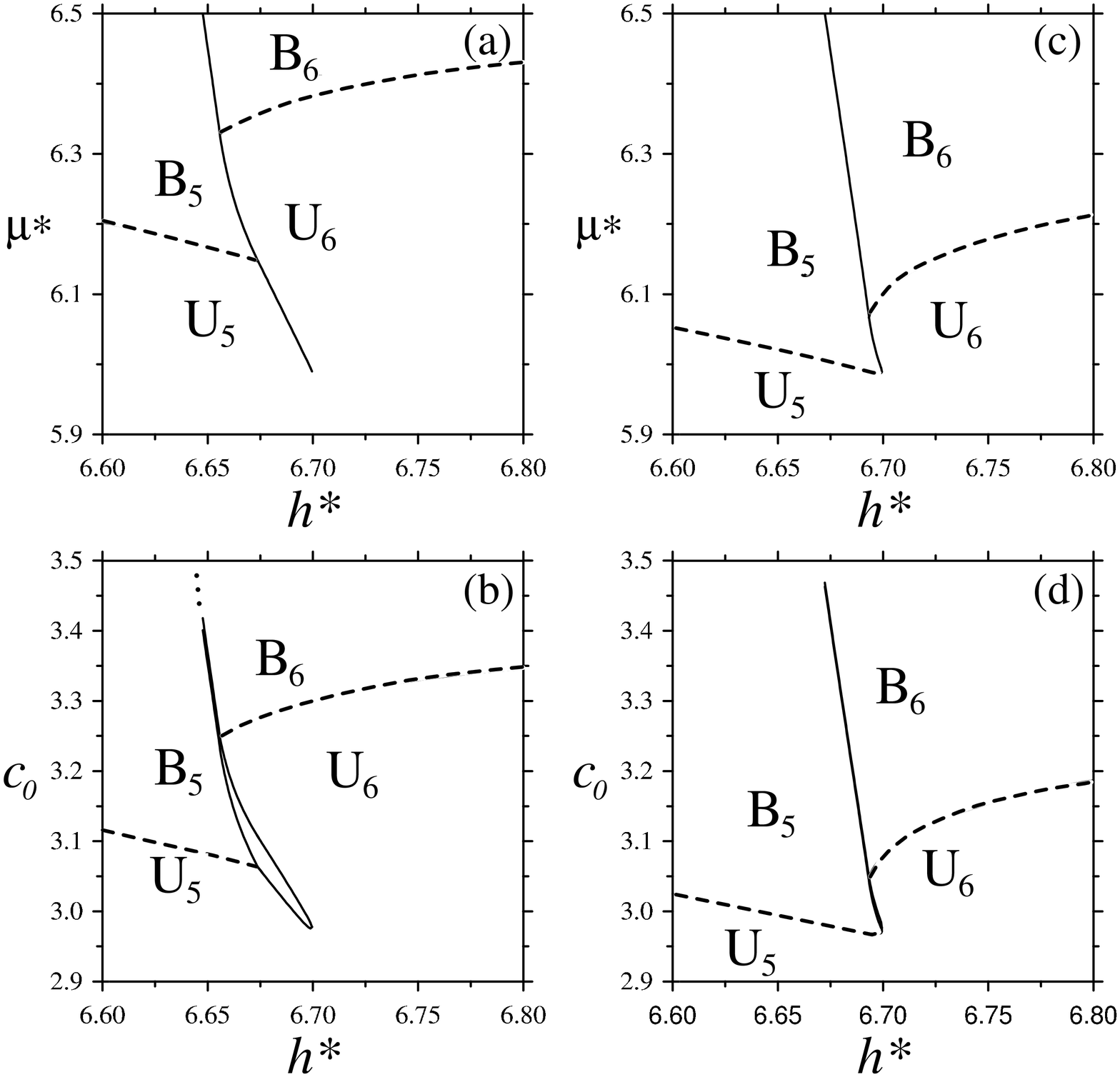}
\caption{\label{fig7} Phase boundaries of layering (continuous curves) and U-B (dashed curves) 
phase transitions.  Upper panels (a) and (c): $\mu^*$-$h^*$ plane. Lower panels (b) and 
(d): $c_0$-$h^*$ plane. U$_5$ and U$_6$ 
denote uniaxial phases with 5 and 6 layers, while B$_5$ and B$_6$ correspond to biaxial phases.
Left panels (a) and (b): $\kappa=12.7$; right panels (c) and (d): $\kappa=13.3$.}
\end{figure*}

The interplay between biaxiality and layering transition is revealed in Fig. \ref{fig7}, 
where we demonstrate the destabilising effect of increasing the cross-sectional aspect ratio on the 
biaxial ordering for the 5-6 layering transition. Not surprisingly, the uniaxial-biaxial
boundary curve moves down in chemical potential due to the increased molecular biaxiality. 
But, in addition, there is a reduction in mean density gap at the layering transition on
increasing $\kappa$ from $12.7$ to $13.3$. Also interesting is that the uniaxial phase
is reentrant in a narrow range of pore widths as $\mu$ or $c_0$ is increased, with 
a phase sequence U$_5\to$B$_5\to$U$_6\to$B$_6$.
The reentrant phenomenon can be seen most clearly in the case $\kappa=12.7$. Finally, as can be 
observed in Fig. \ref{fig7}(c), the U$_5$-U$_6$ transition disappears by increasing the 
aspect ratio.

\begin{figure}[h]
\center\includegraphics[width=3.0in]{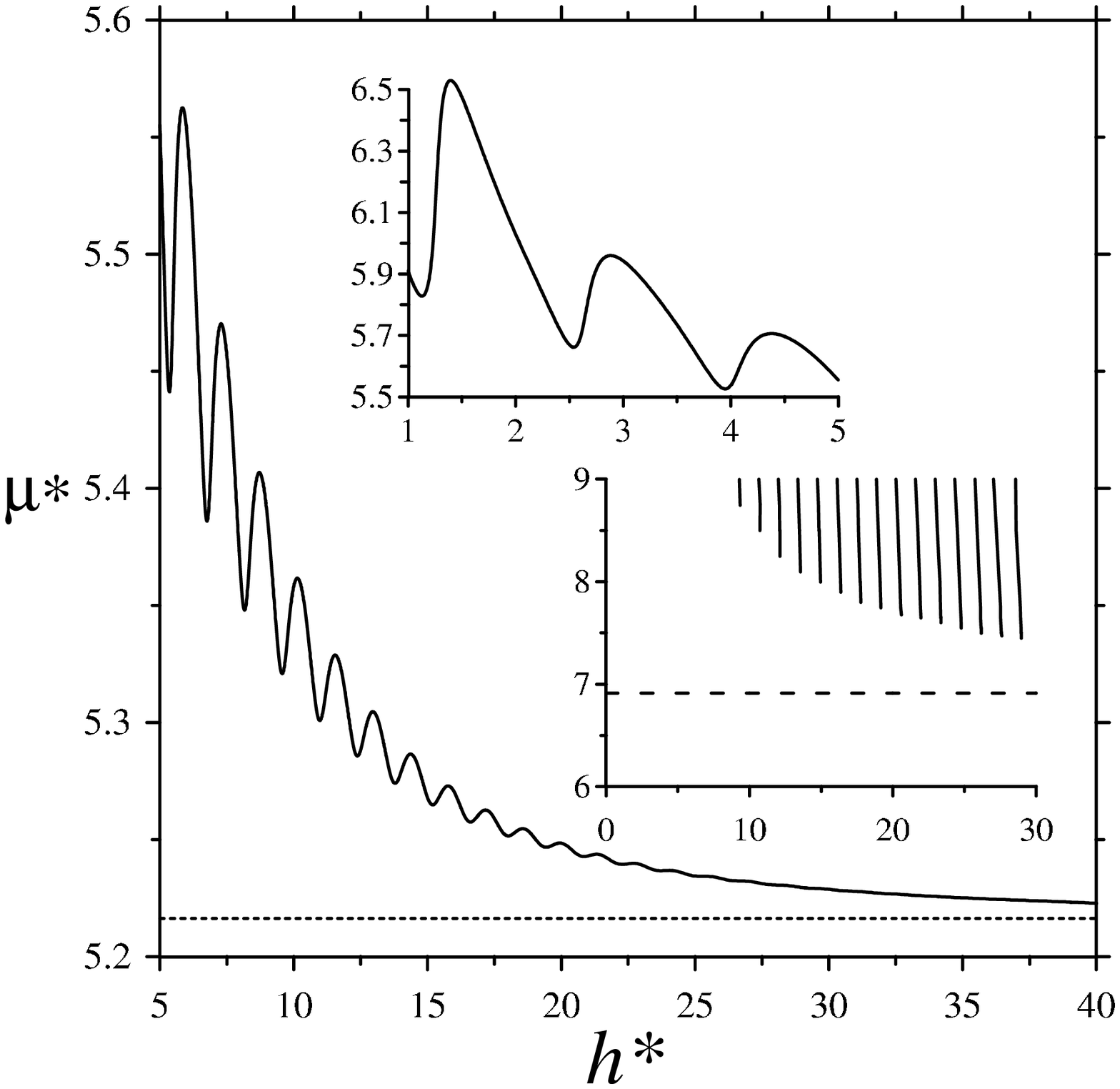}
\caption{\label{fig8} Phase diagram in the $\mu^*-h^*$ plane of the U-B transition for hard cuboids of 
aspect ratio $\kappa=17$. Upper inset shows detail of the same transition 
for small pore widths. Lower inset shows layering transitions ocurring well above 
the U-B transition; horizontal dashed line corresponds to the bulk N$_{\rm{B}}$-S$_{\rm{B}}$.}
\end{figure}

The phase diagram for the $\kappa=17$ fluid is shown in Fig. \ref{fig8}. The U-B phase 
transition curve $\mu_{\rm UB}(h)$ shows damped oscillatory behaviour, and in this case 
converges to the bulk N-N$_{\rm B}$ value with increasing pore width. 
The destabilization effect of the walls on the U-B phase transition can be seen very clearly 
in this case. Since the bulk fluid now
has a large region of N$_{\rm B}$ stability, and the layering transition critical points
converge to $\mu_{\tiny\hbox{N}_{\rm B}\hbox{S}_{\rm B}}>\mu_{\tiny\hbox{NN}_{\rm B}}$
(see inset of Fig. \ref{fig8}),  
the U-B boundary does not intersect the layering transition curves, which now
involve two biaxial phases and end in lower critical 
points. 

\begin{figure}[h]
\center\includegraphics[width=3.00in]{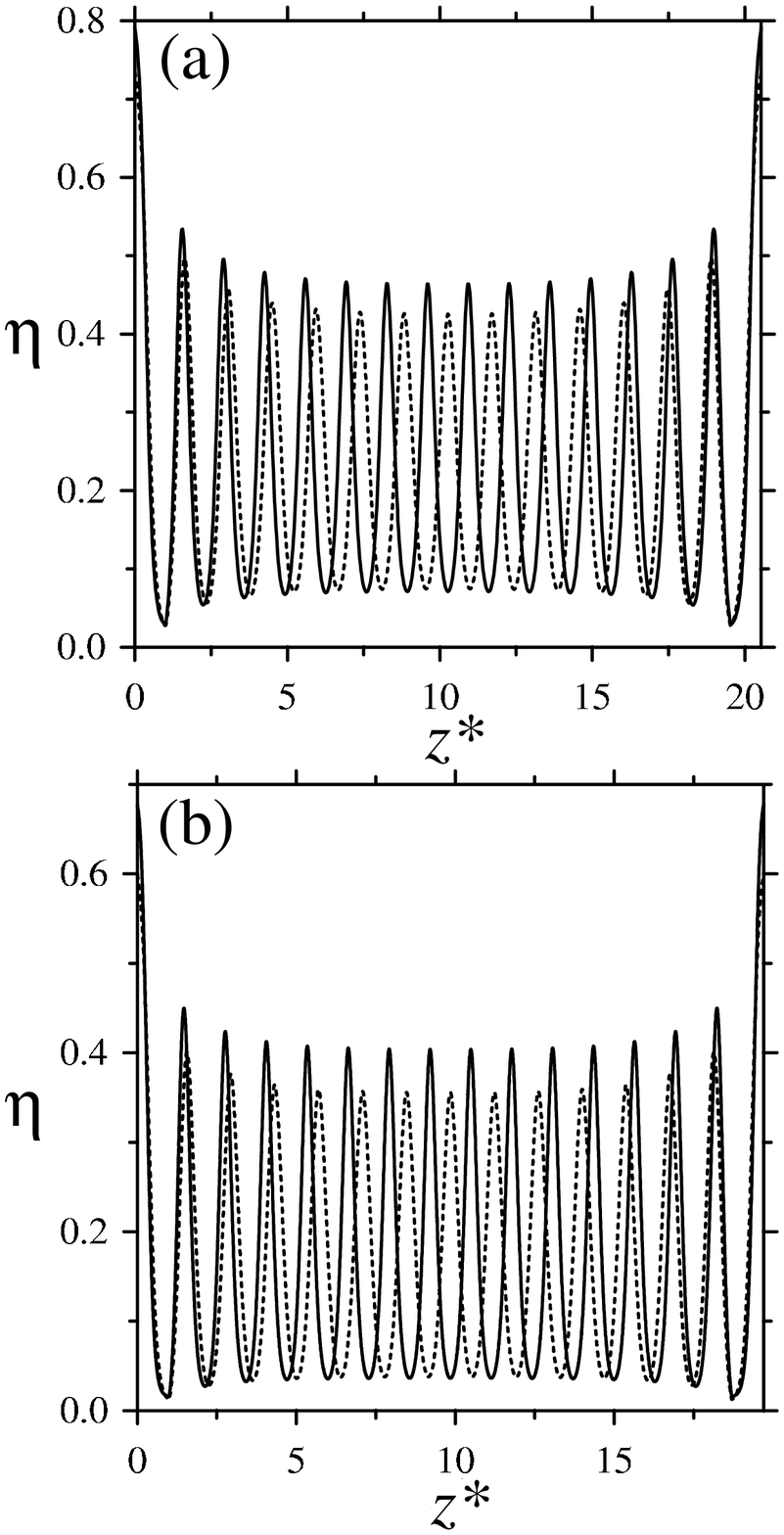}
\caption{\label{fig9} Comparison of the U$_{15}$-U$_{16}$ layering transition density profiles 
from Onsager and FMT theories for $\kappa=13$ and $\beta \mu'=\beta\mu+\log{v_0}=3.5$. (a): 
Onsager theory. (b): FMT theory. 
The dimensionaless local density is defined as $\eta(z)=\rho(z)v_0$, with $v_0=\sigma_1\sigma_2L$.}
\end{figure}

\begin{figure}[h]
\center\includegraphics[width=3.00in]{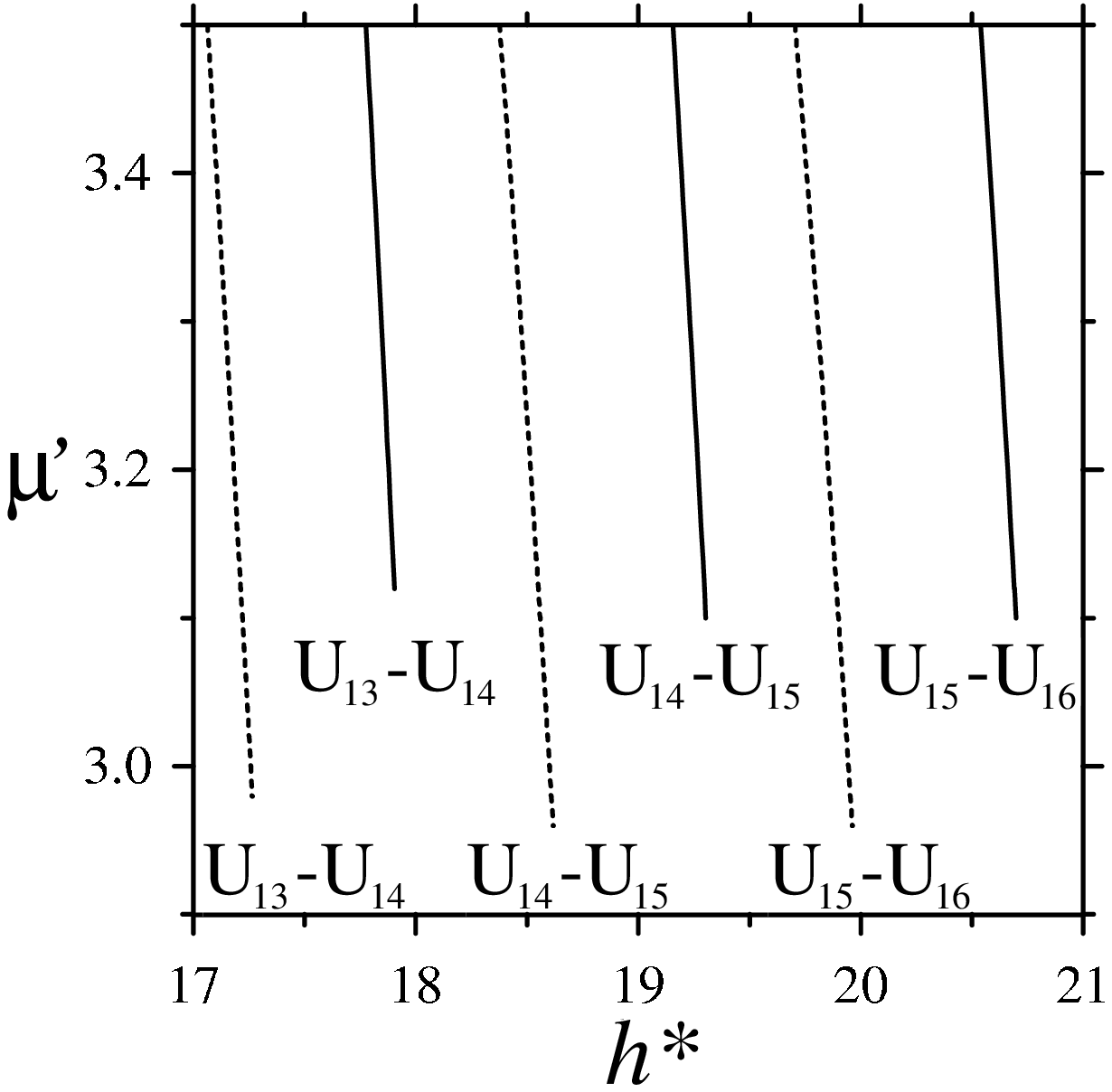}
\caption{\label{fig10} Comparison of the locations of a few layering transitions
in the $\mu^{\prime}$-$h$ phase diagram for uniaxial confined structures. Continuous
curves: Onsager theory. Dashed curves: FMT. $\kappa=13$. A redefined chemical
potential $\mu^{\prime}=\mu^{*}+\log{v_0}$ was used.}
\end{figure}

\begin{figure}[h]
\center\includegraphics[width=3.00in]{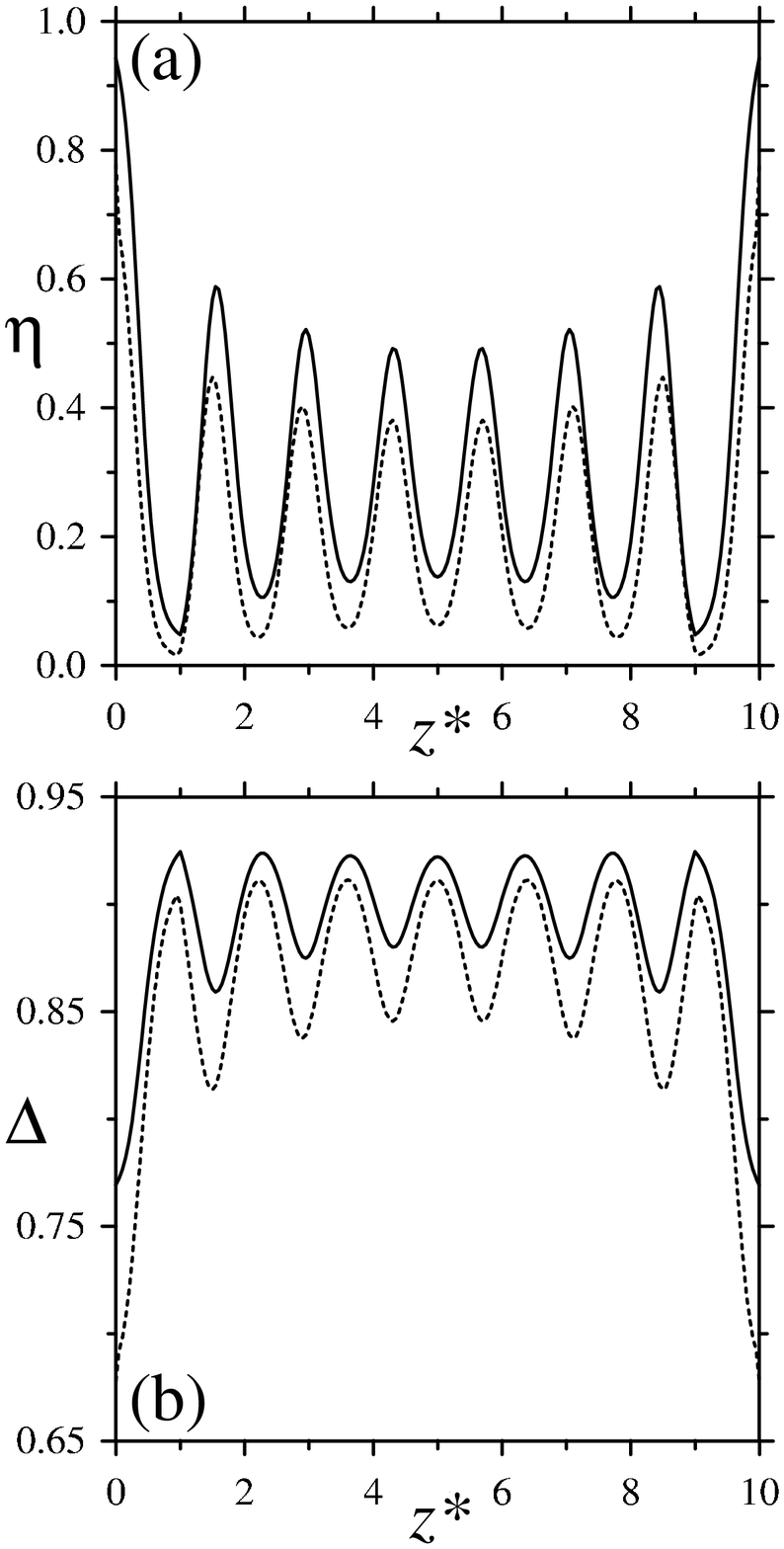}
\caption{\label{fig11} Comparison of density and biaxial order parameter profiles
from Onsager and FMT theories for $\kappa=17$ and $\mu^{\prime}=4.5$ (see caption of Fig. \ref{fig9}).
(a) Density profiles $\eta(z)=\rho(z)v_0$. (b) Biaxial order parameter profile
$\Delta(z)$. Continuous curves: Onsager theory. Dashed curves: FMT.}
\end{figure}

Finally, we make some remarks about the reliability of Onsager theory for confined systems. 
It is true that the second-order virial approximation seems to be a crude approximation for dense 
non-uniform fluids. To assess the validity of Onsager theory, we have made a comparison 
between the predictions of our theory and those of our previous FMT approximation \cite{Us}
formulated for the inhomogeneous fluid. In Fig. \ref{fig9} we focus on a particular (15-16) 
layering transition; the density profiles of the two coexisting
structures at the layering transition, as obtained from Onsager and FMT theories, and at a given 
value of $\mu$, are shown. In Fig. \ref{fig10}
the locations of a few layering transitions in the $\mu$-$h$ phase diagrams are compared,
while Fig. \ref{fig11} shows density and biaxial order parameter
profiles for a given chemical potential. Overall, the agreement between the two theories 
is quite fair. This can be understood if we take into account that for high aspect ratios 
the Onsager theory usually gives correct phase behaviour. 
The oscillatory structure predicted by Onsager theory is overestimated, while
the width of the density peaks is a bit wider than it should be. These discrepancies are due to 
the poor treatment of correlations in Onsager theory. The location of the 
layering transitions is shifted in pore width. However, Onsager theory can be
trusted as far as qualitative behaviours are concerned, and the predicted phase diagram
is expected to be qualitatively correct.

\section{Conclusions}
\label{Sec4}

We have examined the effect of confinement on the phase behaviour of a fluid of
parallel biaxial hard cuboids. Using Onsager theory, four phases are observed in the bulk limit, 
namely uniaxial nematic, uniaxial smectic, biaxial nematic and biaxial smectic. For  
shape anisotropy with $\kappa<15.304$, the sequence 
N$\to$S$\to$S$_{\rm B}$ is observed with increasing chemical potential, 
while the uniaxial nematic phase transforms directly into the biaxial nematic phase for 
$\kappa>15.304$, i.e. in this case
the phase sequence is N$\to$N$_{\rm B}\hspace{-0.1cm}\to$S$_{\rm B}$. 

In the fluid confined into a slit pore only two phases occur: one has uniaxial structure, 
while the other is biaxial. The N-S transition cannot survive, but layering transitions 
do exist and uniaxial-biaxial phase transitions do occur. Layering transitions are associated
to commensuration effects between a structure with an integer number of layers and
a finite pore width, while biaxial ordering 
is the result of excluded-volume interactions taking place between the cross sections of the 
biaxial particles. Interestingly, layering transitions can be destabilized with respect to
an increasing particle biaxiality, characterised by a shape parameter $\kappa$, which means 
that particles with a circular cross sections are the best candidates 
for observing layering transitions. 

Layering transitions between uniaxial-uniaxial,
uniaxial-biaxial, or biaxial-biaxial, structures can occur at high chemical potentials,
and reentrant behaviour of the uniaxial phase may occur for increasing chemical potential 
and for some values of pore width. Confinement does not encourage the formation of biaxial order,
since the uniaxial-biaxial phase boundary moves to higher chemical potential
(density) for decreasing pore width. Therefore, for real fluids that can be modelled by
this particular hard wall/fluid model system, such as colloidal suspensions of anisotropic
particles, experimental detection of the biaxial nematic phase 
will be more favourable in bulk than in confined geometry.

{Our results can be compared to those obtained earlier by van Roij 
et al. \cite{vanRoij,vanRoij1}, who used an Onsager model in the Zwanzig (restricted orientations) 
approximation on hard parallelepipeds of dimensions $L\times\sigma\times\sigma$, with $L\gg\sigma$, to 
analyse surface and capillary behaviours.
In the Zwanzig model particles can only point along the $\hat{\bm x}$, $\hat{\bm y}$ or
$\hat{\bm z}$ directions, and the fluid can accordingly be considered as a mixture of three
species. If we take $L=\sigma_1$, our model can be related to the one in Refs. \cite{vanRoij,vanRoij1},
with the important difference that, in our case, the species perpendicular to the walls is missing, 
so that the isotropic phase cannot occur and there is no capillary nematization/isotropization in
our model (by contrast, particles can freely rotate in the $xy$ plane, and our particle volume is 
finite, so that layered interfacial profiles can be obtained). As a consequence, 
our U-B transition occurs inside the nematic region, whereas van Roij et al.
obtain it below capillary nematization (i.e. in a nematic film adsorbed on the walls), and consequently
there is no dependence of the U-B transition line with pore width (in contrast with our strong
oscillatory dependence, cf. Figs. \ref{fig5} and \ref{fig8}). Our work is a step forward in
the study of confined fluids of hard particles, in that layered smectic phases, not contemplated
before, are included, the relationship between biaxiality and layering is ascertained, and the
dependence of these phenomena with particle cross-sectional aspect ratio is analysed.
}

{Finally,} 
our study shows that Onsager theory can be applied for bulk and confined studies even in 
very ordered phases like the uniaxial or the biaxial smectic phases. Comparison of the 
results of Onsager with FMT theories indicates that, by including higher-order 
correlations into the theory, only quantitative, rather than qualitative, improvement 
can be achieved. We think that further studies should be performed in order to get a 
deeper understanding of the delicate interplay between the surface and biaxial interactions 
on the stability of layering, capillary and orientational transitions. It is not clear how 
the scenario would change if planar, instead of homeotropic, anchoring is favoured, or if
the orientational entropy is included by studying fluids of freely rotating biaxial rods.

Support from Comunidad Aut\'onoma de Madrid (grants S-0505/ESP-0299 and NANOFLUID), Spain, and grants 
FIS2007-65869-C03-01, FIS2008-05865-C02-02 and MOSAICO of the Ministerio de Educaci\'on y Ciencia of 
Spain are acknowledged.

\appendix

\section{Bifurcation analysis} 

In this section we provide some analytical results for the N-S and
N-N$_{\rm B}$ phase transitions of the bulk system, and also for the location of the
U-B transition in the confined fluid. We also give an explanation for the peculiar behaviour of the
biaxial order parameter near the S-S$_B$ transition.

\subsection{N-S bifurcation}
\label{N-S}

The bifurcation analysis of the N-S phase transition of parallel hard cylinders has been 
discussed in earlier work \cite{Mulder}. Here we only present a short overview of the 
bifurcation analysis and show the resulting equations for the bulk bifurcation density 
and wavenumber of the present biaxial hard particles. The starting point is 
the free energy of the smectic phase, given by Eqns. (\ref{6}), for a weak periodic
perturbation of amplitude $\epsilon$:
\begin{eqnarray}
\rho(z)=\rho_0\left(1+\epsilon\cos{qz}\right),
\end{eqnarray}
where $q$ is the wavenumber. Substitution in Eqns. (\ref{6}) provides the free-energy
difference $\Delta F$ between smectic and nematic phases to lowest order in $\epsilon$:
\begin{eqnarray}
\frac{\beta\Delta F}{V}=\frac{1}{4}\epsilon^2\left[\rho_0+2\rho_0^2LA_{\rm ex}^{(0)}
j_0(qL)\right],
\end{eqnarray}
where $j_0(x)$ is a spherical Bessel function. At the N-S bifurcation point, 
\begin{eqnarray}
\Delta F(\rho_0,q)=0,\hspace{0.6cm}\frac{\partial\Delta F}{\partial q}=0,
\end{eqnarray}
and the transition density results in
\begin{eqnarray}
\rho_0^{\rm NS}=-\left[2B_2j_0(q_{\rm NS}^*L)\right]^{-1},
\end{eqnarray}
where $\pi<q_{\rm NS}^*L<3\pi/2$ is the smectic wave number at bifurcation, which can be obtained
as a solution of the trascendental equation $\tan{q_{\rm NS}L}=q_{\rm NS}L$.
The dimensionless density $c_0^{\rm NS}=B_2\rho_0^{\rm NS}$ 
is insensitive to the particular particle model used.

\subsection{N-N$_{\rm B}$ bifurcation}
\label{N-NB}

Similar to the N-S bifurcation analysis, here we determine the free-energy cost
associated with an infinitesimal biaxial ordering in the uniform nematic phase. 
The density distribution 
\begin{eqnarray}
\rho(\varphi)=\frac{\rho_0}{\pi}\left(1+\epsilon\cos{2\varphi}\right),
\end{eqnarray}
gives
\begin{eqnarray}
\frac{\beta\Delta F}{V}=\frac{1}{4}\epsilon^2\left[\rho_0+2\rho_0^2LA_{\rm ex}^{(1)}\right],
\end{eqnarray}
with $A_{\rm ex}^{(1)}=2\pi^{-1}\int_0^{\pi}d\varphi\cos{2\varphi}A_{\rm ex}(0,\varphi)$.
Again the condition $\Delta F=0$ provides a N-N$_{\rm b}$ bifurcation density, given by
\begin{eqnarray}
\rho_0^{\rm NN_{\rm B}}=-\left[2LA_{\rm ex}^{(1)}\right]^{-1},
\end{eqnarray}

\subsection{U-B bifurcation in confined fluid}
\label{U-B}

Unlike the bulk nematic phase, the confined system is nonuniform, and the uniaxial local 
density profile has to be determined from Eqn. (\ref{8}) using a value for the chemical
potential corresponding to the bifurcation point, which is not known a priori. Using Eqn. 
(\ref{8}) it is easy to prove, by integrating out the $\varphi$ dependence, that 
the uniaxial density distribution is the solution of the equation
\begin{eqnarray}
\rho(z)=\exp{\left[\displaystyle\beta\mu-A_{\rm ex}^{(0)}\int_{z-a(z)}^{z+b(z)} dz_1 \rho(z_1)\right]}.
\label{A8}
\end{eqnarray}
The biaxial perturbation is again proportional to the function $\cos{2\varphi}$, but its 
amplitude must depend on the position in the pore. Therefore we use the following ansatz for 
the perturbation:
\begin{eqnarray}
\rho(z,\varphi)=\frac{\rho(z)}{\pi}\left[1+\epsilon F(z)\cos{2\varphi}\right].
\label{A9}
\end{eqnarray}
Substitution into Eqns. (\ref{6}) provides a free-energy difference between the uniaxial
and the biaxial confined structures, which is quadratic in $\epsilon$. The coefficient
must be zero at the bifurcation point, which provides the equation
\begin{eqnarray}
&&\hspace{-0.4cm}\int_0^h dz\frac{f^2(z)}{\rho(z)}\nonumber\\\nonumber\\&&+\frac{A_{\rm ex}^{(1)}}{2}\int_0^h
dz_1 f(z_1) \int_{z_1-a(z_1)}^{z_1+b(z_1)} dz_2 f(z_2)=0,
\label{A10}
\end{eqnarray}
where $f(z)=F(z)\rho(z)$.
This equation is still not suitable for the determination of $f(z)$, which requires
minimizing the perturbed free energy with respect to $f(z)$. This corresponds to the 
functional derivative of (\ref{A10}) with respect to $f(z)$ being zero, i.e. 
\begin{eqnarray}
f(z)=-\frac{1}{2}A_{\rm ex}^{(1)}\rho(z)\int_{z-a(z)}^{z+b(z)}dz_1f(z_1).
\label{A11}
\end{eqnarray}
The simultaneous solution of Eqns. (\ref{A8}), (\ref{A10}) and (\ref{A11}) gives the density 
profile $\rho(z)$, chemical potential $\mu$ and perturbation function $f(z)$ at the 
uniaxial-biaxial phase transition. From the set of U-B bifurcation equations one can get 
analytical results in the case $0<h<L$. This interval corresponds to a very narrow pore, 
where only one monolayer of particles can fit in the pore. The upper and lower bounds of the 
integral in (\ref{A8}) become $0$ and $h$, respectively,  i.e. the integral will give the 
same value for any position in the pore. This means that the local density is constant 
in such a narrow pore. Its value depends on the chemical potential and is given by 
\begin{eqnarray}
\rho(z)=\rho_0=e^{\displaystyle\beta\mu-A_{\rm ex}^{(0)}h\rho_0}.
\label{A12}
\end{eqnarray}
Similar to the local density, (\ref{A11}) predicts a constant value $f(z)=f_0$ and a density
at bifurcation given by
\begin{eqnarray}
\rho_0^{\rm UB}=-\frac{2}{hA_{\rm ex}^{(1)}}.
\label{A13}
\end{eqnarray}
Note that (\ref{A13}) gives the same bifurcation equation as (\ref{A11}).
$\mu_{\rm UB}$ can be obtained from (\ref{A12}) using (\ref{A13}).

\subsection{Smectic phase biaxiality}
\label{S-biaxiality}
This section is devoted to showing how the biaxial ordering of particles develops near the 
S-S$_{\rm B}$ transition for the particular case where the S phase is highly ordered. We explicitly show  
the peculiar behaviour of the order parameter profile as a periodic function peaked 
around the midpoint between adjacent smectic layers. 

Substitution of Eqn. (\ref{A9}) into the definition of the 
biaxial order parameter, i.e.
\begin{eqnarray}
\Delta(z)=\rho(z)^{-1}\int_0^{\pi} \rho(z,\varphi)\cos{2\varphi}d\varphi 
\end{eqnarray}
provides the result $\Delta(z)=F(z)/2$. Now insertion of the definition 
$f(z)=F(z)\rho(z)=2\Delta(z)\rho(z)$ into Eqn. (\ref{A11}) allows us to 
find 
\begin{eqnarray}
\Delta(z)=-\frac{1}{2}A_{\rm{ex}}^{(1)}\int_{z-L}^{z+L}dz_1\rho(z_1)\Delta(z_1),
\label{A13a}
\end{eqnarray} 
which can be viewed as an integral equation for the biaxial order parameter 
profile $\Delta(z)$ near the S-S$_{\rm{B}}$ transition. The first derivative of 
Eqn. (\ref{A13a}) with respect to $z$ gives
\begin{eqnarray}
&&\hspace*{-0.5cm}\Delta'(z)=-\frac{A^{(1)}_{\rm{ex}}}{2}
\left[\rho(z+L)\Delta(z+L)-\rho(z-L)\Delta(z-L)\right].\nonumber\\
&&\hspace*{-0.5cm}
\end{eqnarray}
Due to the periodicity of the order parameter profile, the function 
$\Delta(z\pm L)$ is periodic with period $d$:
\begin{eqnarray}
\Delta(z\pm L)=\Delta\left[z\pm (L-d)\right].
\end{eqnarray}
Its Taylor expansion about $z$, up to first order, gives
\begin{eqnarray}
\Delta(z\pm L)\approx\Delta(z)\pm\Delta'(z)(L-d).
\end{eqnarray}
For the particular case of a highly ordered smectic phase (with $d\sim L$),
we can make the approximation $\Delta(z\pm L)\approx\Delta(z)$, where use has been made
of the fact that $\Delta'(z)$ has the same order of 
magnitude as $\Delta(z)$ near the bifurcation point. Note that 
this approximation is not justified for the function $\rho(z\pm L)$ in general, because we 
have assumed that the smectic phase is highly ordered and thus that the first 
derivative of the density profile $\rho'(z)$ can reach high values for some values of $z$. 
Making all these approximations, we find the following diferential equation for the order parameter:  
\begin{eqnarray}
\Delta'(z)=-\frac{A_{\rm{ex}}^{(1)}}{2} \Delta(z)\left[\rho(z+L)-\rho(z-L)\right].
\end{eqnarray}
The solution to this equation is 
\begin{eqnarray}
\Delta(z)&=&\Delta(0)\exp{\left[\Psi(z)-\Psi(0)\right]},\label{A14a}\\\nonumber\\
\Psi(z)&=&-\frac{A^{(1)}_{\rm{ex}}}{2}\int_{z-L}^{z+L}\rho(z')dz'.\label{A14}
\end{eqnarray}
Assuming that the periodic density profile $\rho(z)$ is peaked at
$z=z_k\equiv kd$ ($k\in\mathbb{Z}$), it is easy to show that the maxima of the 
function $\Psi(z)$ given by (\ref{A14}) are located at $z=z_k+d/2$, resulting in 
an order-parameter profile $\Delta(z)$ in antiphase with respect 
to the density profile $\rho(z)$ [see Eqns. (\ref{A14a}) and (\ref{A14})]. The present result 
is valid for any hard non-local interaction with a range of twice the particle length.

\end{document}